# Should my Blockchain Learn to Drive?
# A Study of Hyperledger Fabric

Jeeta Ann Chacko
chacko@in.tum.de
Technical University of Munich

Ruben Mayer
ruben.mayer@uni-bayreuth.de
University of Bayreuth

Hans-Arno Jacobsen
jacobsen@eecg.toronto.edu
University of Toronto

**Abstract**

Similar to other transaction processing frameworks, blockchain systems need to be dynamically reconfigured to adapt to varying workloads and changes in network conditions. However, achieving optimal reconfiguration is particularly challenging due to the complexity of the blockchain stack, which has diverse configurable parameters. This paper explores the concept of self-driving blockchains, which have the potential to predict workload changes and reconfigure themselves for optimal performance without human intervention. We compare and contrast our discussions with existing research on databases and highlight aspects unique to blockchains. We identify specific parameters and components in Hyperledger Fabric, a popular permissioned blockchain system, that are suitable for autonomous adaptation and offer potential solutions for the challenges involved. Further, we implement three demonstrative locally autonomous systems, each targeting a different layer of the blockchain stack, and conduct experiments to understand the feasibility of our findings. Our experiments indicate up to 11% improvement in success throughput and a 30% decrease in latency, making this a significant step towards implementing a fully autonomous blockchain system in the future.

## 1 Introduction

The increasing complexity of transaction processing systems, such as databases, led to the development of self-adaptive [31, 33], self-tuning [11, 75], and self-managing systems [20, 41]. The eventual objective of this discipline is to create a self-driving system that can autonomously predict workload and network changes, and reconfigure itself to optimal performance without human intervention. Research is currently underway to achieve this goal [39, 40, 45–47, 57, 58, 76]. Since blockchains have evolved to support complex transactions using smart contracts, they are now categorized as transaction processing systems [63]. This raises the question of whether self-driving blockchains are feasible.

Blockchain systems, like other transaction processing systems, require dynamic reconfiguration to cope with changes in workloads and network conditions [9, 43, 69]. However, optimal reconfiguration is particularly challenging due to the complexity of the blockchain stack, which has diverse configurable parameters [10, 43]. These include administrative policies, database definitions, consensus protocols, ledger settings, and smart contract design, among others [9, 42, 62, 72]. Therefore, domain expertise is essential for achieving optimal performance in blockchain systems. However, this expertise comes at a high cost, with the estimated maintenance cost of a blockchain application being up to 25% of the total development cost [36, 67, 68]. As long as blockchains remain complex and expensive, enterprises will hesitate to adopt this platform for their use cases.

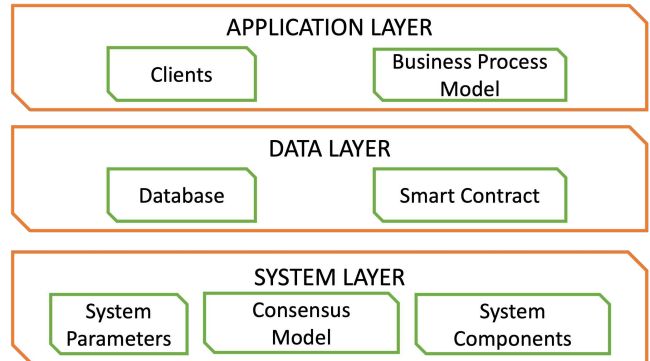

Figure 1: Adaptable Features

Consequently, a self-driving blockchain that can eliminate human involvement is desirable. Some initial steps in this direction have been taken through the development of self-adaptive and auto-tuning blockchain systems [43, 72, 73]. However, the existing systems are either not completely autonomous and hence require human intervention or are confined to tuning a single aspect of the blockchain stack. Unlike the database community, the blockchain literature has yet to explore the concept of a comprehensive self-driving blockchain extensively. Self-driving systems can either be created from scratch [58, 76] or by building upon existing systems [46]. The development of over 1000 different blockchain systems worldwide in a relatively short time span has made it challenging for enterprises to choose the ideal blockchain for their applications [26]. Rather than augmenting this problem, we focus on exploring self-driving possibilities in existing systems. Further, stakeholders of established systems may be reluctant to switch to a new blockchain platform and integrating autonomous capabilities into existing systems helps to avoid this requirement. Adapting the configuration settings of public blockchains may lead to hard forks as not all network participants may accept the changes [62]. In contrast, permissioned blockchains, which are more commonly used by enterprises, would welcome such changes if they can improve the overall performance. Hyperledger Fabric is one of the most popular permissioned blockchains with over 50 enterprise partners [25, 60].

This paper delves into the opportunities for self-drive in Hyperledger Fabric. Our investigation involves identifying *adaptable features*, i.e., the parameters and components of the blockchain stack that can be dynamically tuned to improve performance, as illustrated in Figure 1. We address specific challenges and offer a blockchain perspective on the topic of self-driving systems. Furthermore, we implement a prototype and conduct experiments to evaluate the feasibility of our findings.

Our contributions can be summarized as follows:

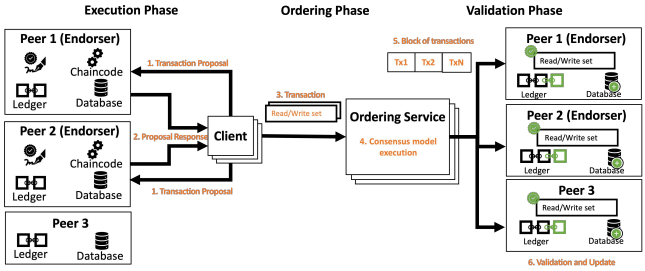

Figure 2: Transaction Flow in Hyperledger Fabric

(1) We scrutinize the various dynamic aspects of the blockchain environment, such as network and workload evolution, to determine the necessity of self-driving blockchains. We compare and contrast our findings with existing research on databases and highlight aspects unique to blockchains. Our goal is to encourage the blockchain research community to contribute more to the development of self-driving blockchains.

(2) We conduct a thorough analysis of the entire system stack of Hyperledger Fabric and identify specific parameters and components, which we call *adaptable features*, that are suitable for autonomous adaptation. Our study also highlights the challenges in making these adaptable features autonomous and provides potential solutions. By doing so, we encourage the blockchain community to explore self-driving opportunities within existing systems instead of creating a new blockchain from scratch for each new use case.

(3) We evaluate our findings by setting up three demonstrative autonomous systems, each targeting a different level in the blockchain stack. Our results indicate up to 11% improvement in success throughput and 30% decrease in latency. This is a significant first step towards implementing a fully autonomous system in the future. To the best of our knowledge, this is the first comprehensive discussion and evaluation of self-driving blockchains.

(4) The implementation of our systems, experimental workloads, and trained models are made available as open source. The research community can use these resources as a foundation to conduct further investigations with other adaptable features and machine learning strategies. This will facilitate further discussions on self-driving blockchains.

## 2 Hyperledger Fabric

One of the most popular open-source permissioned blockchain systems, Fabric [2], was established under the Linux Foundation. Fabric is unique in that it allows for smart contracts to be created using general-purpose languages. This gives clients the ability to submit transactions to a decentralized network, where control is shared among multiple entities instead of a single trusted entity. All potential functions that can be executed within a transaction are defined in a smart contract, referred to as chaincode. The system maintains a distributed, versioned key-value store known as the world state.

Each key in the network has a version number that gets updated with every write. The platform maintains a comprehensive history of all transactions, both successful and failed, through a distributed ledger that groups transactions into blocks. This distributed ledger, along with the world state, gets replicated on a set of distributed nodes called peers that are registered on the Fabric network. The peers receive blocks of transactions from an ordering service that guarantees ordered delivery and validate each transaction independently before updating their copy of the world state and the ledger.

Endorsers are a subset of peers that not only validate transactions but also endorse and execute the transactions. The number of endorsements required for a transaction to be accepted as valid is defined by an endorsement policy. Furthermore, peers are grouped into organizations, which typically correspond to real organizations or branches of an enterprise. These organizations play an important role in the endorsement policy as such policies can specify the number of endorsements required from each organization's respective peers. In Fabric, the transaction flow consists of three main phases: execution, ordering, and validation. This process is commonly known as the Execute-Order-Validate (E-O-V) model, and it is visualized in Figure 2.

*Execution Phase.* When clients need to update data stored in the blockchain, they initiate a transaction proposal to the endorsers. This transaction can include multiple requests for reads and writes to one or more keys in the world state. The endorsers simulate the transaction's execution on the current world state, generating a read/write list for each key involved in the transaction. They then send a response to the client, which includes their signature and the read/write set. During the transaction flow, the client collects responses from the endorsing peers and forwards them to the ordering service nodes.

*Ordering Phase.* The ordering service uses a consensus protocol, such as Raft [55] or BFT [5], to order transactions received from the clients. A transaction block is created based on three conditions: if a fixed duration of time has elapsed (block timeout), if a fixed number of transactions has been received (maximum message count), or if the total size of transactions has reached a fixed limit (maximum preferred bytes). The ordered block of transactions is then sent to all peers.

*Validation Phase.* When the ordering service sends a block of transactions, each peer validates every transaction in the block independently. Each peer verifies if a sufficient number of valid endorsing peer signatures, based on the endorsement policy, have been collected (Validation System Chaincode (VSCC) validation). After that, the peer checks if the version of each key in the read set of each transaction is equal to the version of the same key in the current world state (Multi-Version Concurrency Control (MVCC) validation). If both VSCC and MVCC validation checks pass, the write sets of the transactions are applied to the world state. However, if any of the validation checks fail, the client is informed that the transaction has been aborted, and the world state does not change. Once the validation is complete, the validated block containing both aborted and committed transactions is added to the ledger. The status of every transaction, whether committed or aborted, is logged for future reference.

## 3 Need for self-driving blockchains

In this section, we explore the reasons that necessitate the implementation of self-driving blockchains. Our analysis draws upon insights from the database literature while also shedding light on the unique concepts that are specific to blockchains.

## 3.1 Workload Evolution

Blockchains, similar to other transaction processing systems, have to handle frequently varying workloads. Depending on the application, workload variation may follow typical diurnal patterns, such as higher transactions during the day than night or spike patterns, such as sudden influx during Christmas in a supply chain management scenario [30, 57]. Apart from these familiar patterns, since blockchains are geographically distributed, the dynamic addition of participants from various time zones and their corresponding transactions can change the overall workload structure [51].

Unlike other transaction processing systems, blockchains also need to process administrative transactions apart from application-related transactions. For example, a *configuration transaction* needs to be executed when the system is reconfigured, such as changing the block size or integrating a new peer [2, 70]. Such transactions also follow the complete transaction lifecycle. Users may also trigger historical queries that read the complete or parts of the blockchain ledger to confirm the validity of their own transactions [66]. Such transactions are highly time-consuming.

When considering Fabric specifically, it has an optimistic concurrency control model where transactions can fail due to data dependency [64]. In such cases, the client may resend failed transactions immediately or later, depending on the business process logic of the enterprise. Additionally, Fabric's FIFO ordering strategy can result in situations where one type of transaction overwhelms the system, blocking all other transactions [30].

In summary, blockchains handle heterogeneous workloads that follow unpredictable arrival patterns and require the processing of additional administrative transactions, making them a complex system to manage. The optimal configuration of various system parameters, network components, smart contracts, database models, consensus algorithms, and business process models greatly depends on the workload [9, 10, 69]. Currently, only static auto-tuning systems are available for users to determine the best settings for their specific type of workload [43]. However, to use such systems, it is crucial to obtain appropriate representative workloads. This proves to be particularly challenging for permissioned blockchains, which are primarily used for enterprise purposes, since private organizations are hesitant to reveal their workloads. As a result, a self-driving blockchain that has the ability to monitor its evolving workload and dynamically adjusts itself to the ideal settings can be a promising solution.

## 3.2 Network Evolution

Scaling in blockchains is highly heterogeneous and dynamic [12, 65]. In permissioned blockchains, the different system components are mapped to physical entities in an enterprise. For example, the peer nodes of a Fabric network are grouped into administrative units called *organizations* that typically correspond to the physical organizations or branches of a company [2]. As a result, real-world administrative activities of enterprises, such as expanding their global reach or acquiring other organizations, necessitate adapting the blockchain network. Moreover, specific network components, such as endorsers and orderers, which have additional privileges, such as executing the smart contract and ordering the transactions, may be reassigned to different geographical locations depending on enterprise management changes in the physical world. System parameters must be adjusted to support such network scaling. For instance, if the number of ordering nodes is too high, communication costs increase, and performance is negatively affected. In such cases, Fabric recommends dynamically redesigning the network into subsets called channels and deploying separate ordering node sets per channel [16].

Additionally, even without any changes in the network components, the blockchain ledger grows perpetually over time. Therefore, the network must be constantly monitored and adjusted accordingly to avoid any bottlenecks. For example, Fabric recommends increasing the resources whenever the CPU, memory, or disk space usage reaches 70%, as high resource utilization significantly impacts performance [16]. However, since there are multiple distributed system components (peers, endorsers, orderers, clients, database, ledger), monitoring and identifying the bottleneck is challenging. Further, given the decentralized nature of blockchains, the participants need to reach consensus before taking scaling decisions. Consequently, a self-driving blockchain that constantly monitors the evolving network, identifies bottlenecks, triggers the consensus mechanism and automatically adapts the network configuration would be highly beneficial.

## 3.3 Performance and Fairness

Self-driving systems are designed to achieve multiple goals, including optimal performance in terms of throughput and latency. This is particularly true for self-driving transaction processing systems such as databases as well as blockchains [39, 40, 43, 45–47, 57, 58, 72, 73, 76]. The ability to sustain adequate throughput despite workload or network changes without (or with minimal) human intervention is the ultimate goal.

Another important objective for blockchains is fairness. Since blockchains lack centralized entities, transactions are generally processed in a first-in, first-out (FIFO) order, which may result in geographically closer and resource-intensive clients being able to commit more transactions [51]. In enterprise scenarios where many participants have equal administrative rights, such unequal representation of their transactions on the ledger may raise trust issues [7, 30]. To address this problem, a self-driving blockchain that identifies dominant clients, controls their transaction admission rates, and ensures fairness in the network is urgently needed. Such a system would ensure that all participants are equally represented on the ledger, thereby promoting trust and transparency.

## 4 Self-Driving Opportunities & Challenges

Since we have established the need for a self-driving blockchain in Section 3, our main objective now is to determine the feasibility of such a system. To achieve this, we must ascertain which components and configuration parameters require dynamic adaptation when the workload or network evolves. Some system configuration parameters that have an impact on the performance have been identified in the literature [43]. However, a majority of these parameters cannot be tuned without restarting the blockchain network, which is not possible in a live network. In contrast, we will examine specific features that can be adjusted without requiring a network restart and are thereby suitable for designing a self-driving blockchain. Additionally, we will solely concentrate on features that can be modified at runtime without significant alterations to the original system's architecture. We analyze the complete blockchain stack using experimentation and literature review to identify such adaptable features. In this section, we explain each of these features and discuss the challenges of dynamically adapting them. The experimental setup, workloads and metrics definition for Figures 3-5 can be found in Section 6.

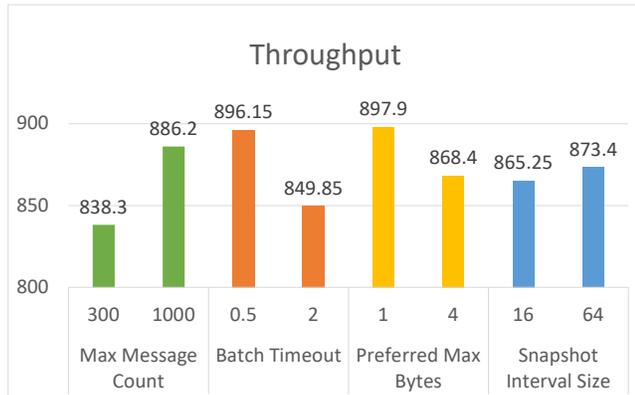

Figure 3: Impact of configuration parameters on performance

## 4.1 System Layer

We conducted an experimental analysis of various dynamically configurable system parameters to determine the adaptable features of Fabric at the system layer. The main results that highlight the impact of the parameters on overall throughput for a send rate of 1000 TPS are illustrated in Figure 3.

### 4.1.1 Max Message Count

The *max message count* refers to the maximum number of transactions allowed in a block. The ideal value for *max message count* varies based on the workload and the performance metric being considered. For instance, in our experiments, higher values of *max message count* significantly improve the overall throughput (cf. Figure 3). The literature recommends setting the *max message count* to match the incoming transaction rate of the workload [9, 16]. However, research has also shown that if the transaction rate is below a system's throughput saturation point, lower values for *max message count* are optimal [69]. Since all the experiments in the literature, as well as ours, are conducted with different workloads under different network conditions, we cannot derive a consistent relation between *max message count* and performance. Therefore, the *max message count* can be identified as an adaptable feature that needs to be dynamically adjusted based on the evolving workload and network.

### 4.1.2 Batch Timeout

The *batch timeout* is the maximum timeout after which a block is created with the currently available transactions. We observe that tuning this parameter has an impact on the performance. For example, when *batch timeout* is set to 2 seconds, which is the default value, it negatively impacts the overall throughput (cf. Figure 3). The official recommendation from Fabric is to set this value to *max message count* divided by the transaction rate [16]. Therefore, since *max message count* is an adaptable feature and the transaction rate is variable, *batch timeout* must be considered as an adaptable feature.

### 4.1.3 Preferred Max Bytes

The *preferred max bytes* refers to the maximum size (in MB) of all the transactions allowed in a block. Our findings indicate that the throughput is impacted by *preferred max bytes* (cf. Figure 3). Fabric recommends setting *preferred max bytes* to *max message count* multiplied by the average transaction size [16]. Since the optimal value of *preferred max bytes* depends on the incoming workload, it can be classified as an adaptable feature.

### 4.1.4 Snapshot Interval Size

The *snapshot interval size* parameter of the consensus protocol in Fabric (Raft) defines the number of bytes per which a snapshot of the log is taken. This is the only dynamically tunable parameter for Raft. While creating snapshots at regular intervals reduces disk space usage, it can be an expensive process [24]. Therefore, dynamically tuning this parameter based on the incoming load and disk usage can be helpful. We observe that an increase in *snapshot interval size* has a slight impact on the throughput (cf. Figure 3). Due to its dependency on the incoming workload, *snapshot interval size* can be identified as an adaptable feature

*Lessons Learnt*: Our experiments help to understand the impact of individual system configuration variables on the performance of Fabric. However, manually deriving the best combination of values for these parameters would be costly and brittle. This highlights the need for a self-driving blockchain. Self-driving systems generally adopt machine learning strategies, which involve exploring multiple values until the system learns the ideal setting. However, this process may face several obstacles, one of which is transaction queueing caused by the block size. In our experiments, we observed that creating multiple small blocks may overwhelm Fabric's ordering service when the transaction rate is high, causing the network to hang. Further, in our experiments, a very low *snapshot interval size* for a high transaction rate also led to system hang-ups. Hence, it is crucial to carefully choose the values of system parameters and transaction rates during the learning phase to prevent such issues in the Fabric network.

## 4.2 Data Layer

The data layer comprises the blockchain ledger, the smart contracts, and the database. In Fabric, the ledger and database cannot be dynamically reconfigured, so the focus is on optimizing the smart contract performance. Smart contracts play a pivotal role in the functioning of blockchains. To optimize their performance, various strategies are employed, such as delta writes, smart contract partitioning, and primary key redefinition [10]. However, the effectiveness of these strategies depends on the workload. For instance, *delta* writes can convert update transactions that increment a variable into write-only transactions, reducing transaction dependency failures in update workloads. However, this optimization strategy can negatively impact the performance of compute workloads. Therefore, adapting the smart contract according to the workload would be useful. To examine this, we created a smart contract that allows for a value to be incremented as an update transaction or as a write-only transaction using two different function implementations: *vanilla* and *delta*. We evaluate the performance impact of both implementations using an update workload and a compute workload. More details about the smart contract and the workloads can be found in Section 6.2.

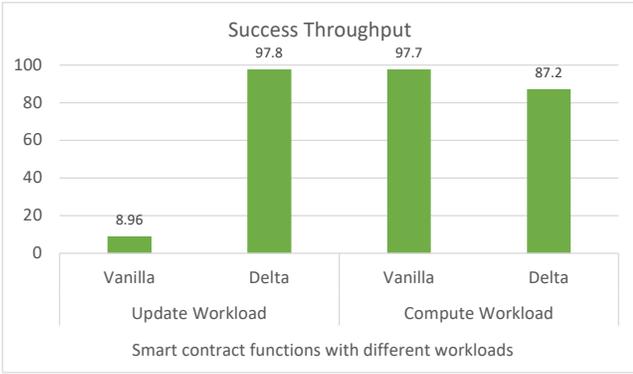

**Figure 4: Impact of smart contract on performance**

Our experimental results shown in Figure 4 indicate that the use of the *delta* implementation results in a significant improvement in success throughput for update workloads. However, for compute workloads the success throughput decreases with the use of the *delta* implementation. Further, over a longer duration, we observed that Fabric is unable to sustain the high latency resulting from the use of the *delta* implementation with compute workloads (not shown in Figure 4). In Fabric, smart contracts can be upgraded in real-time, making them an adaptable feature that can be customized to meet the specific requirements of different workloads.

*Lessons Learnt:* In enterprise scenarios, smart contracts are often defined as automated executions of contractual agreements between entities in the real world [23, 28]. As a result, frequent upgrades to smart contracts are generally discouraged. Our solution to this issue is to include multiple implementations of the logic within the same smart contract and selectively invoke the desired implementation from the client side based on the varying workload. This approach enables greater flexibility and reduces the need for frequent upgrades. Further, based on our experiments, we conclude that for a self-driving blockchain system, sufficient duration must be given for each learning step to correctly understand the effect of the adaptable features.

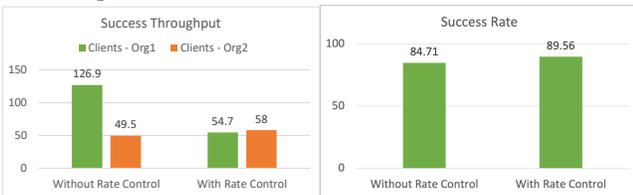

**Figure 5: Impact of rate control on performance**

### 4.3 Application Layer

The clients are the main components of the application layer, and fairness among the clients is crucial for building users' trust in a blockchain system. However, ensuring fairness is difficult due to the diversity in the geographical locations of the clients relative to endorsing peers and the resources available to the clients. There are instances where high transaction rates from a few clients congest the system, leaving the other clients with very low throughput [30]. Fabric's optimistic concurrency control strategy also follows a first-come-first-serve model, which can cause transactions from slower clients to fail more frequently. Monitoring such bottlenecks and dynamically adapting client transaction send rates is essential to ensure fairness. Therefore, the admission rate of clients is a potential adaptable feature.

We conducted an experiment that simulates an unfair distribution of transactions, where clients of one organization (clients-Org1) have a higher transaction send rate than the clients of another organization (clients-Org2). Further, the transactions generated by clients-Org1 have key conflicts with the transactions generated by clients-Org2. More details about the workloads can be found in Section 6.3. From our experimental results shown in Figure 5, it is evident that rate control can ensure fairness, but it comes at the cost of degraded overall success throughput. However, we can see in Figure 5 that the total success rate improves. This experiment clearly demonstrates the tradeoff between fairness and overall performance.

*Lessons Learnt:* The main challenge in this direction is to design an optimal fairness strategy. If the transaction rate of faster clients is significantly restricted, it can have a severe impact on overall performance. Moreover, such restrictions may not even lead to a corresponding increase in the success rate of slower clients. Therefore, it is crucial to consider both fairness and overall performance when designing a self-driving system.

## 5 Self-driving Blockchain System Design

To demonstrate self-driving capabilities, we integrated a prediction system and a monitoring system with the Fabric network. The monitoring system extracts performance metrics from the client network: overall throughput (in TPS), success throughput (in TPS), average latency (in seconds), number of successful transactions per client and success rate per client. The prediction system controls various adaptable features of the Fabric network and is explained in detail in this section. Figure 6 provides a visualization of the architecture of our system.

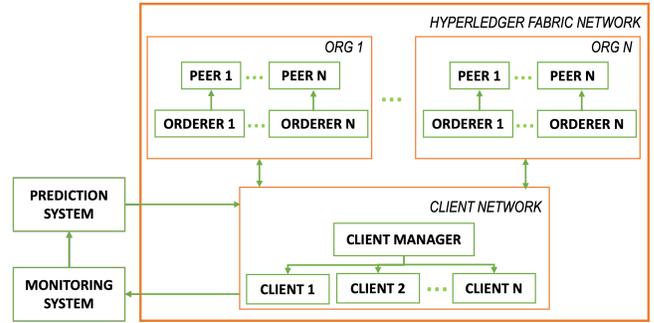

**Figure 6: Self-driving Blockchain System**

Our prediction system uses a reinforcement learning-based approach to autonomously reconfigure the adaptable features. Reinforcement learning [38] is a machine learning strategy that involves the agent applying an action, observing the consequence of its action on the environment, receiving a reward based on the consequences, and altering its actions over time to maximize the reward. It is mainly used in scenarios where training data is not initially available, and it is popularly used in self-driving systems [43, 75, 76]. Permissioned blockchains are mainly used by enterprises that are generally hesitant to share their workloads and ledger contents due to privacy concerns. Therefore, there is a lack of publicly available training data, which makes reinforcement learning a suitable approach for our use case. We implement our prediction system using a reinforcement learning library called Deep Q Network (DQN) provided by Stablebaselines [21].

For a reinforcement learning agent (RL agent), three parameters need to be defined. The *state* is the environment that needs to be observed by the agent. The *action space* is a discrete or continuous set of actions that the agent is allowed to take. The *reward function* is a quantification of the reward that an agent receives based on the effect of its action on the state. The reward function depends on the performance expectations of the blockchain system user, which is often defined in a service level agreement. The definition for each of these parameters varies based on our target adaptable feature and is explained in the upcoming sections.

During each step of the learning process, the prediction system communicates with the Fabric network through its clients. It notifies the clients of any required changes in configuration, smart contract logic, or admission rate. To ensure that these changes are implemented and have an impact on the network, the learning process is paused for a while before moving on to the next step. The monitoring system provides performance information, which the prediction system uses to update its *state* and calculate its *reward*.

As a tangible initial step towards realizing a self-driving block chain, we demonstrate the effect of dynamically modifying the identified adaptable features. At the system level, we demonstrate parameter tuning, where the *max message count*, *batch timeout*, *preferred max bytes* and *snapshot interval size* are dynamically adapted. At the data level, we demonstrate the self-drive capability by dynamically adapting the smart contract implementation. Finally, client admission rate is tuned dynamically to understand the potential for self-driving at the application level. For our experiments, we designed three locally autonomous systems based on reinforcement learning, which can pave the way to a completely self-driving system in the future.

## 5.1 RL Agent for Parameter Tuning

This section describes our autonomous parameter tuning mechanism. The RL agent of the prediction system needs to dynamically learn the optimal values of four configuration parameters – *max message count* ($M_C$), *preferred max bytes* ($P_B$), *batch timeout* ($B_T$), and *snapshot interval size* ($S_I$), Therefore, the action space ($A_S$) of the RL agent consists of the set of values that each of these parameters can adopt. We chose these values based on intuition and previous research. The default value of *max message count* is 500. Studies show that it is optimal if the *max message count* matches the incoming transaction rate [9, 16], which in our experiments is 300, 500, and 1000 TPS. The default value of *preferred max bytes* and *snapshot interval size* is 2 MB and 16 MB, and in our experiments, values less than the default led to a significant decrease in the performance (even such that the system hangs). Therefore, we chose the default value as well as two values higher than the default. The default value of *batch timeout* is 2s, and increasing this value significantly increases the latency if none of the other parameters related to block size are fulfilled by the incoming transactions. Therefore, we chose the default value as well as two values lower than the default. The action space ($A_S$) for the RL agent is then the cross-product of all possible values for all four parameters.

$$M_C = [300, 500, 1000]$$
$$P_B = [2, 4, 16]$$
$$B_T = [0.5, 1, 2]$$
$$S_I = [16, 32, 64]$$
$$A_S = [M_C \times P_B \times B_T \times S_I]$$

The reward function ($R_W$) of an RL agent depends on the performance expectations of the blockchain system user. In this experimental setting, we assume that maximizing the throughput ($T$) of the Fabric network is the primary goal. The average transaction send rate ($SR$) of the clients vary with time. Therefore, we need to consider the throughput relative to the send rate.

$$R_W = \frac{T}{SR}$$

Similarly, since the impact of the agent's action is measured by the change in the throughput relative to the send rate, the *state* ($S_T$) or environment that the RL agent needs to observe is determined by these values.

$$S_T = [T, SR]$$

At every learning step, the RL agent picks an action ($A$) from the action space randomly or based on previous experience.

$$A = [m_c, p_b, b_t, s_i]$$
$$where\ m_c \in M_C, p_b \in P_B,\ b_t \in B_T,\ s_i \in S_I$$

The chosen action is packaged as a *configuration transaction* and sent to the Fabric network via its clients. After this transaction is endorsed, ordered and validated by the peers in the network, the action is applied, i.e., the configuration parameters are updated, and the RL agent moves to the next learning step.

## 5.2 RL Agent for Smart Contract Adaptation

In this section, we discuss the mechanism for autonomous smart contract adaptation. The RL agent needs to dynamically learn the optimal smart contract implementation – *vanilla* and *delta*, which we represent as 0 and 1 in the action space($A_S$) for the RL agent.

$$A_S = [0, 1]$$

Adapting the smart contract aims to improve the success throughput. As a result, the reward function ($R_W$) of the RL agent is defined by the success throughput ($S_{UT}$) relative to the send rate ($SR$).

$$R_W = \frac{S_{UT}}{SR}$$

Similarly, since the impact of the agent's action is measured by the change in the success throughput relative to the send rate, the *state* ($S_T$) is determined by these values.

$$S_T = [S_{UT}, SR]$$

The RL agent picks an action ($A$) which is either the *vanilla-update* or the *delta-update* implementation at every learning step.

$$A = [a_s]\ where\ a_s \in A_S$$

We created a new configuration file for the Fabric client to define the smart contract implementation. The client's invocation of the smart contract is dependent on this configuration file, which is updated by the RL agent with the chosen action.

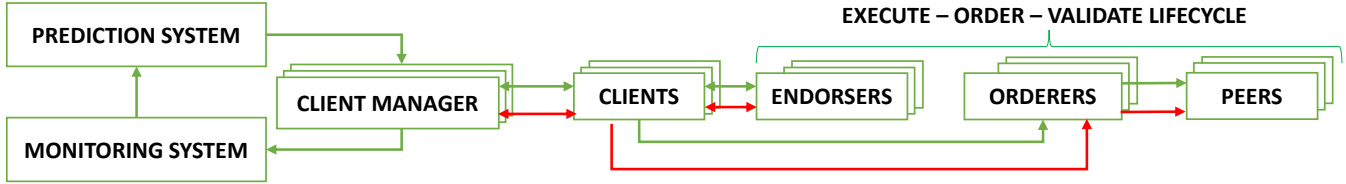

Figure 7: Life cycle of transactions generated in a Fabric network with (green) and without (red) the prediction system

## 5.3 RL Agent for Admission Rate Tuning

In this section, we delve into the mechanism of dynamic admission rate tuning with the aim of guaranteeing fairness across all clients. The RL agent needs to dynamically control the send rate of *clients-org1* ($S_R Org_1$) and *clients-org2* ($S_R Org_2$). The send rate cannot be increased, since in a real-world scenario, the clients would only send the required transactions and not create new transactions just to match a given send rate. Therefore, the send rate can either be throttled or kept unchanged. We define a decrease of 40% and 60% of the send rate based on intuition and experimentation. A decrease of less than 40% may not significantly influence the success rate of the other clients, and a decrease of more than 60% could hurt the overall throughput. Therefore, the send rates of *clients-org1* and *clients-org2* can either remain unchanged, decrease by 40% or decrease by 60%.

$$S_R Org_1 = [unchanged, decrease_{40}, decrease_{60}]$$
$$S_R Org_2 = [unchanged, decrease_{40}, decrease_{60}]$$
$$A_S = [S_R Org_1 \times S_R Org_2]$$

The main goal of this experiment is to ensure that the number of successful transactions generated by *clients-org1* ($S_{uc}T_r Org_1$) and the number of successful transactions generated by *clients-org2* ($S_{uc}T_r Org_2$) are as close to equal as possible. We use the Jain's fairness index ($\mathcal{J}_{fi}$) to quantitatively define fairness in the range (0,1], where higher values indicate a fairer distribution [37]. This index ($\mathcal{J}_{fi}$) is used to define the reward function.

$$\mathcal{J}_{fi} = \frac{(S_{uc}T_r Org_1 + S_{uc}T_r Org_2)^2}{2 * (S_{uc}T_r Org_1^2 + S_{uc}T_r Org_2^2)}$$

$$R_W = \mathcal{J}_{fi}$$

The impact of the agent's action is also measured by the change in the success throughput relative to the send rate and the fairness measure. Therefore, the *state* ($S_T$) is determined by these values.

$$S_T = [S_{UT}, SR, \mathcal{J}_{fi}]$$

The RL agent chooses an action ($A$) at every learning step which either decreases or maintains the admission rate of the clients.

$$A = [sr_1, sr_2] \text{ where } sr_1 \in S_R Org_1, \ sr_2 \in S_R Org_2$$

We have developed a new configuration file for the Fabric client, which defines the transaction rate per organization. This configuration file is used to adapt the transaction rates of each organization. This file is updated by the RL agent with the chosen action at every learning step.

## 5.4 Decentralization Aspects

Our prediction system is designed as a centralized system that is independent from the blockchain network. This allows for easier implementation of updates, monitoring of the training process, and adherence to regulations and standards. Further, training systems typically result in high resource utilization over time. Since our prediction system is independent of the Fabric network, it can be deployed on a separate node with high resource allocation without encroaching on the resources required by the blockchain system. Additionally, such an independent prediction system can be readily replaced with different learning algorithms based on user requirements.

However, blockchains rely heavily on decentralized trust, which is an essential feature of their operation. Therefore, a self-driving blockchain also needs to maintain this decentralized nature. Though our prediction system is centralized the inherent decentralization properties of Fabric is maintained. The changes proposed by the prediction system are sent to a client manager, which then generates a *configuration transaction* for parameter tuning or a transaction with specific parameters to adapt the smart contract logic. The client manager also defines the transaction admission rate of each client. The client manager is a simulation of an automated business process execution system often used in enterprise scenarios to manage the execution of an application [71]. In a production-level setup, there could even be more than one client manager, and the prediction system would communicate the proposed changes to all of them. However, all transactions generated by the client manager undergo the complete execute-order-validate lifecycle defined by the Fabric network, i.e., all network participants need to reach a consensus when any change proposed by the prediction system is applied. As visualized in Figure 7, the lifecycle of transactions with and without the prediction system remains the same. Therefore, the decentralization and security guarantees of Fabric are maintained despite having a centralized prediction system.

## 6 Experimental Setup

We conducted our experiments on four clusters each with 1 master node and 3 worker nodes. All nodes are deployed with 100 GB memory and 30 GB storage. The master nodes have 32 virtual CPUs while the worker nodes have 16 virtual CPUs each. On all clusters, we launched a Fabric network on the workers with four peers (two per organization) and three orderers. Additionally, we deployed ten clients along with a client manager using Hyperledger Caliper, a benchmarking system for Hyperledger Fabric. We ran the prediction system and monitoring system on the master node of all clusters. We developed different workloads and smart contracts for evaluating the three autonomous systems as described in the following.

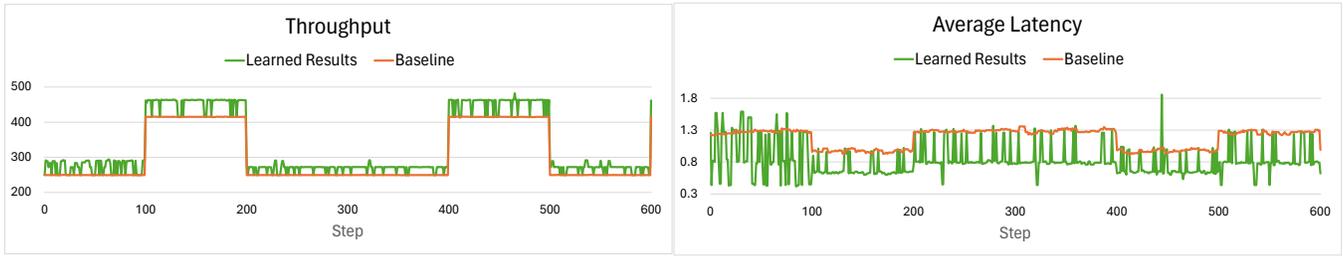

**Figure 8: Update workload with and without autonomous parameter tuning**

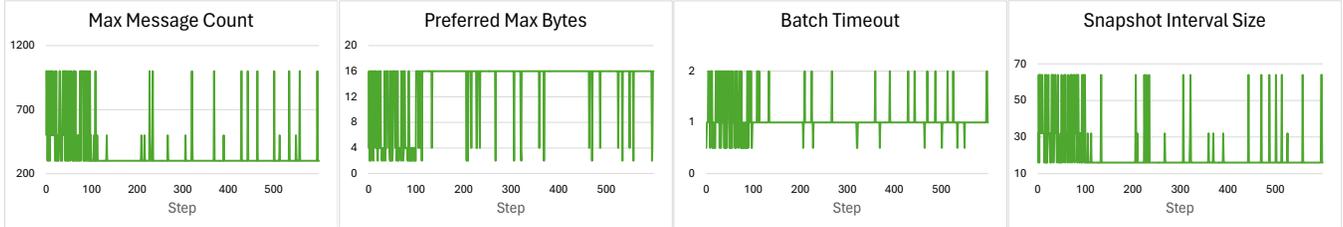

**Figure 9: Values of parameters set by the prediction system for the update workload**

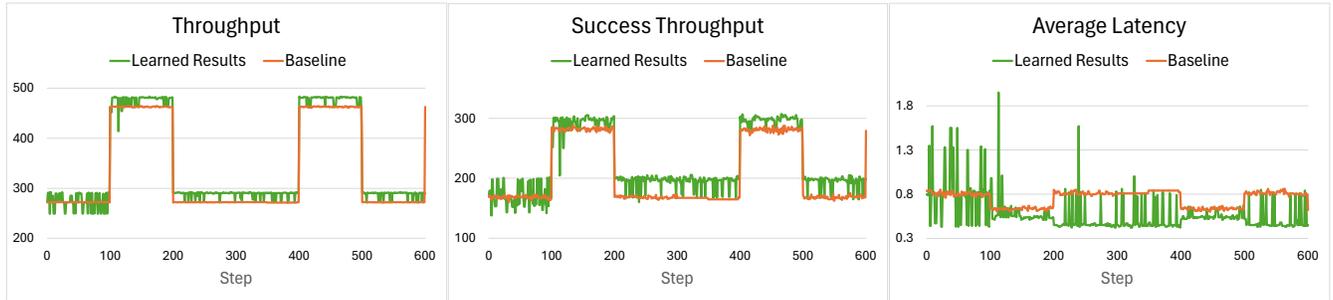

**Figure 10: Skewed update workload with and without autonomous parameter tuning**

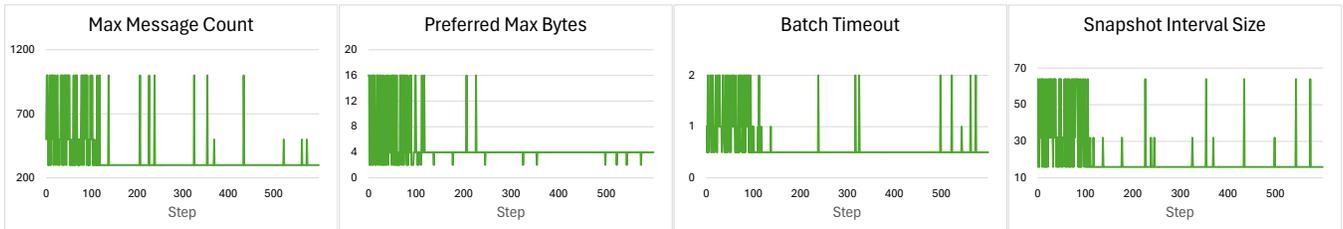

**Figure 11: Values of parameters set by the prediction system for the skewed update workload**

### 6.1 Workloads for Parameter Tuning

We use the *generator* smart contract from the literature [9], which can generate update transactions (single key read and write). Using this smart contract, we generate two different workloads: *update* workloads which have no transaction dependencies and *skewed update* workloads which have multiple transaction dependencies. We initialize the blockchain ledger with 10,000 keys. For the duration of our experiments, the clients' average transaction send rate oscillates between 300 TPS and 500 TPS at regular intervals of 100 steps to simulate an evolving workload. For all the experiments, the baseline is evaluated by setting the default value of 500, 2 MB, 2 seconds, and 16 MB for *max message count, preferred max bytes, batch timeout,* and *snapshot interval size,* respectively.

### 6.2 Workloads for Smart Contract Adaptation

We implemented a *music management* smart contract that includes the two main functions, *PlayMusic* and *CalculateRevenue*, from a music rights management scenario based on the literature [9]. In our database, every entry has a structure consisting of three components: *musicID*, *play-count*, and *total-revenue*. The *PlayMusic* function retrieves the entry for a given *musicID*, increments the value of the *play-count* by one, and writes it back to the database. This implementation of the function, which we call the *vanilla* implementation, results in an *update* transaction. The same function has an alternative implementation, the *delta* implementation, which writes a new entry into the database in the format *musicID-operation-value-transactionID* (for example, M01+1T01). In other words, every increment to the *play-count* creates a new entry in the database, making the transaction *write-only* instead of an *update*.

The *vanilla* implementation of *CalculateRevenue* function reads the value of *play-count* for a given *musicID* from the database. The *delta* implementation needs to aggregate the value of all keys in the database that have the partial format *musicid-operation-value-transactionID* (for example M01+1*) and add this aggregated value to the *play-count*. Such aggregations are time-consuming operations that, depending on the number of keys, can lead to system hang-ups for long-running experiments. Implementing additional smart contract functions that prune the delta keys and regularly invoking these functions as part of the workload can help resolve this issue. However, to reduce the complexity of the smart contract implementation and workload definition in our experimental setup, we simulate this aggregation functionality using a delay of 500ms. This simulation sufficiently demonstrates the effect of the *delta* implementation on performance while preventing disruptions to the experiment. Since our experiments focus on evaluating the prediction system's ability to learn the impact of adaptable features on performance, such a simulation is adequate. After aggregation, the *play-count* is multiplied by a constant and used to update the *total-revenue*.

We initialize the blockchain ledger with 10,000 keys. We generate an *update* workload that only invokes the *PlayMusic* function and a *compute* workload that only invokes the *CalculateRevenue* function. This simulates a real-world scenario where music is frequently played by all system users while the revenue is only occasionally calculated by the artists. The workload oscillates between *update* and *compute* to simulate an evolving workload. The average transaction send rate is 100 TPS. We also run two baselines. Baseline 1 uses only the *delta* implementation, while baseline 2 only uses the *vanilla* implementation.

### 6.3 Workloads for Admission Rate Tuning

We use the same *generator* smart contract and *skewed update* workload used for parameter tuning. We extended the client workload generation logic so that the transaction rates can be adapted per organization. Five of the clients registered to Org1 in the Fabric network send transactions at a rate of 250 TPS, and we call them *clients-org1*. The other five clients, which we call *clients-org2*, send transactions at a rate of 100 TPS. Further, the transactions generated by *clients-org1* have key conflicts with the transactions generated by *clients-org2*. This simulates a real-world scenario mentioned in the literature where one type of transaction floods the network, thereby causing the other transactions to fail [30].

## 7 Results and Observations

We conducted three sets of experiments to demonstrate our three locally autonomous systems. The results and observations from our experiments are described in this section.

### 7.1 Self-driven Parameter Tuning

This section describes our autonomous parameter tuning experiments. Figure 8 shows the overall throughput (which is equal to the success throughput since there are no failures) and average latency of the Fabric network with (learned results) and without (baseline) the prediction system on the *update* workload. The workload's send rate changes at every 100 steps, corresponding to the throughput changes visible in the graph. For the baseline, the values of the four configuration parameters are set to default. The results show that the performance of the network is improved significantly when the prediction system is employed. Specifically, the throughput increases by an average of 7%, and latency decreases by an average of 30%. This validates the positive impact of autonomous parameter tuning. From Figure 9 we observe that the most frequently used values that the prediction system learned over time for the *update* workload are (300, 16 MB, 1 s, 16 MB) for *max message count*, *preferred max bytes*, *batch timeout*, and *snapshot interval size*, respectively.

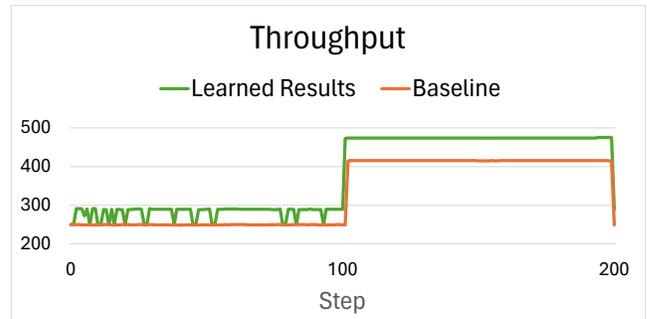

**Figure 12: Autonomous parameter tuning with dynamically decreasing exploration rate**

The experimental results with the *skewed update* workload are visualized in Figure 10. For the baseline, we used the best values (300, 16 MB, 1 s, 16 MB) for the configuration parameters from the previous experiment. We observe that even with the earlier learnt best values, the performance improves with the use of the prediction system. We observe an average of 4% increase in overall throughput, 8% increase in success throughput and 23% decrease in latency. From Figure 11 we observe that the most frequently used values that the prediction system learned over time for the *skewed update* workload are (300, 4 MB, 0.5 s, 16 MB) for *max message count*, *preferred max bytes*, *batch timeout*, and *snapshot interval size*, respectively. We observe that lower values are chosen for *preferred max bytes* and *batch timeout* with the *skewed update* workload. Comparing the baseline latency in Figure 8 and 10 we observe that transactions are processed faster with the *skewed update* workload. This could be due to the different configuration settings or because the presence of transaction failures reduce the number of writes on the database. Since the network is able to process transactions faster, the prediction system chooses to also create blocks faster by reducing the *preferred max bytes* and *batch timeout*.

It is evident that the two workloads have distinct optimal settings and despite our efforts to comprehend the rationale behind these specific selections, manually determining the perfect combination of values would be a daunting task, underscoring the significance of autonomous parameter tuning.

We further conducted an experiment with a dynamically decreasing exploration rate. Exploration rate defines the rate at which the prediction system tries new actions over time to identify the ideal setting. By decreasing this hyperparameter we observe in Figure 12 that a stable throughput that is on an average 11% higher than the baseline can be maintained. However, exploration is generally

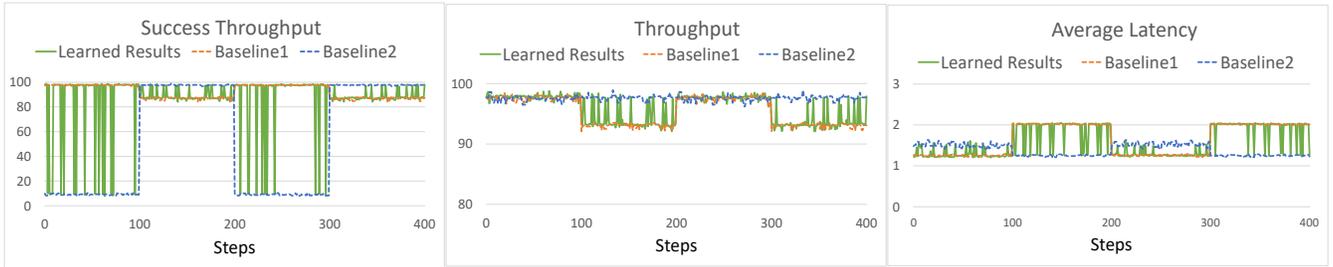
Figure 13: Performance with and without autonomous smart contract adaptation

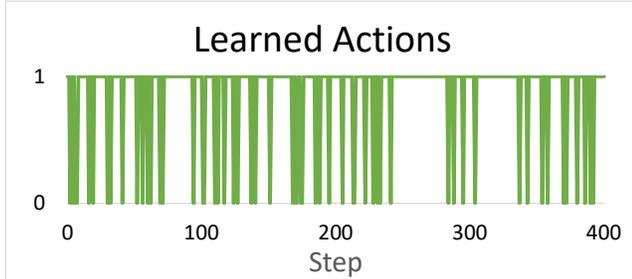
Figure 14: Learned smart contract tuning actions

encouraged in self-driving systems to accommodate unexpected changes that might require further learning.

### 7.2 Self-driven Smart Contract Adaptation

In this section, we discuss the experiments we conducted to enable automatic adaptation of the smart contract to explore the feasibility of autonomous smart contract upgrades. Figure 13 shows the overall throughput, success throughput, and average latency of the Fabric network with (learned results) and without (baseline) the prediction system using the *music management* smart contract. Baseline 1 uses only the *delta* implementation, while baseline 2 only uses the *vanilla* implementation. We initially use the *update* workload and then the workload oscillates between *update* and *compute* workloads at every 100 steps. We observe that Baseline 1 suffers from a very low success throughput of 9 TPS on average with the *update* workload but has high success throughput with the *compute* workload. In contrast, Baseline 2 performs better with the *update* workload and has a decrease in throughput with the *compute* workload.

The prediction system learns the performance impact of the two implementations on both workloads. Choosing the *delta* implementation over *vanilla* gives more than 10 times the success throughput with the *update* workload. At the same time, this choice decreases the success throughput only by an average of 10% with the *compute* workload. Figure 14 shows the actions learned by the prediction system where 0 represents the *vanilla* implementation and 1 represents *delta*. We observe that the prediction system correctly identifies that the *delta* smart contract implementation is a better choice for such an oscillating workload as the gain in performance with the *update* workload far exceeds the loss in performance with the *compute* workload. It chooses the *delta* smart contract implementation around 86% of the time over the *vanilla* implementation.

### 7.3 Self-driven Admission Rate Tuning

In this section, we delve into the experiments we carried out to adjust the rate at which the clients send transactions independently, with the aim of guaranteeing fairness across all clients. Figure 15 shows the Jain's fairness index, success throughput and success rate at each learning step with and without the prediction system.

In Figure 16, the values in the y-axis [0, 0.5, 1] represent the actions [$decrease_{60}$, $unchanged$, $decrease_{40}$] respectively. We observe that over time, the faster clients (*clients-org1*) learn to decrease the transaction send rate, while the slower clients (*clients-org2*) learn to maintain the default transaction send rate. The prediction system learns that decreasing the admission rate of the faster clients can result in a fair distribution of the number of successful transactions between *clients-org1* and *clients-org2* as shown by upto 16% increase in the Jain's fairness index in Figure 15.

We also observe that with the prediction system the slower clients have more successful transactions per learning step than the faster clients (not shown in the figure). However, this tradeoff is acceptable as the overall success rate improves by 6% (Figure 15). In other words, with the prediction system, the faster clients have a lower transaction send rate than the baseline but have a similar success rate to the baseline. The slower clients have a similar transaction send rate and success rate to the baseline, but the number of successful transactions per learning step increases (up to 16% increase, not shown in the figure). Further, there is an average of 30% decrease in overall success throughput (Figure 15), which is expected because the main objective is to ensure fairness between the clients in terms of successful transactions.

### 7.4 Learning Overheads

The prediction system is deployed on a separate node, and therefore, the overheads related to computing and storing the training model do not affect the Fabric components such as peers and orderers. In the parameter tuning experiments, the only interaction with the Fabric network is the execution of a *configuration transaction*. However, the network processes over 15,000 transactions at every step, so a single additional transaction will not have significant overhead. The prediction system updates the client configuration files for the smart contract and admission rate control experiments. The overhead of this file write operation is also insignificant as it happens in a non-blocking manner on separate process threads. Further, the metrics measured in our experiments are inclusive of all overheads that the prediction system might induce.

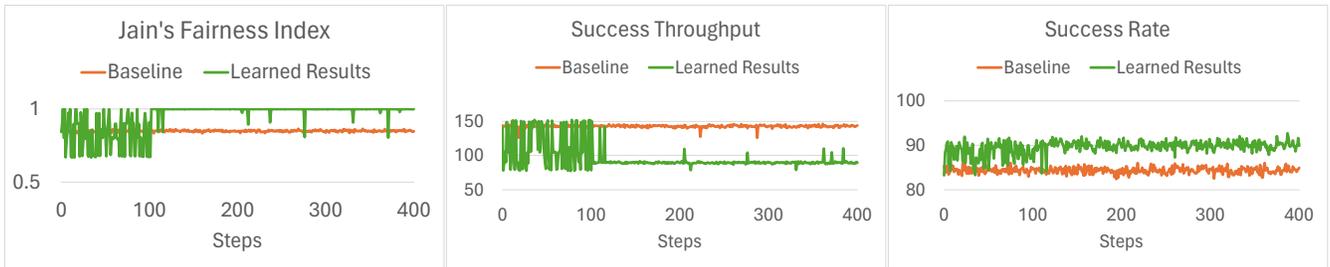
Figure 15: Performance with and without autonomous client admission rate tuning

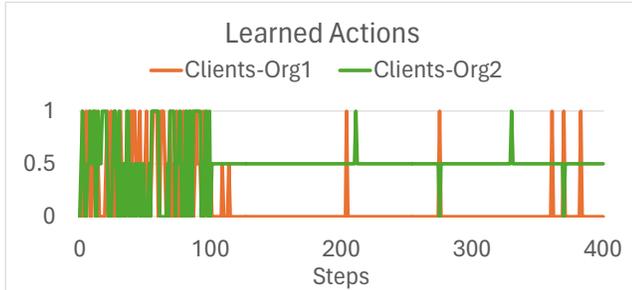
Figure 16: Learned admission rate tuning actions

### 7.5 Key Takeaways

We demonstrated that self-driving capabilities can be incorporated into different levels of the blockchain stack. Our experiments show that the prediction system is able to learn the optimal values for a given workload, leading to improved system performance. Additionally, reducing the exploration rate over time can help the prediction system to converge to these ideal values and produce stable, higher performance. However, this approach could hinder the detection of workload changes that require further learning. It is also important to note that fine-tuning the adaptable features can have varying impacts on each performance metric. For this reason, reward functions must be carefully formulated based on user requirements and considering the potential tradeoffs involved.

## 8 Discussions

We have identified several features that can be dynamically adapted and demonstrated their feasibility. Our findings represent a significant milestone in our quest to develop a completely self-driving blockchain. In this section, we will discuss further potential adaptable features, the transferability of our findings to other blockchain systems, and limitations of our work.

### 8.1 Further Adaptable Features

The adaptable features discussed in this paper so far are versatile and can be applied across various blockchain use cases without significant architectural changes. Additionally, our literature review has identified adaptable features that are more specific to particular use cases and require more extensive architectural modifications. We will discuss these features and their challenges in this section.

#### 8.1.1 BFT Consensus Model

In enterprise use cases, blockchain participants are typically assumed to be non-byzantine, and the network only needs to be crash fault tolerant. However, recently, Fabric has introduced a Byzantine fault-tolerant consensus protocol [5]. The literature includes dynamic leader selection [74] and validator pool size adaptation [42] strategies for BFT protocols that could be included in a self-driving blockchain design. Further, the BFT consensus model in Fabric has 13 configuration parameters that can be tuned dynamically [14]. One such example is the *IncomingMessageBufferSize*, which is the size of the buffer that temporarily stores incoming transactions. Low values for this variable at high transaction rates cause bottlenecks [16]. Therefore, these configuration variables can also be identified as adaptable features.

#### 8.1.2 Network Components

The performance of a Fabric network is greatly influenced by its network components. Research shows that the number and distribution of endorsers and orderers have a significant impact on network performance, and this can vary depending on the incoming workload [9, 10, 15, 16, 69]. Therefore, dynamically scaling these components up or down based on the workload can potentially improve network performance. However, since decentralized trust is a fundamental characteristic of blockchains, the network components are both geographically and administratively distributed among the participating entities. Geographically closer endorsers and orderers ensure faster transaction processing, and therefore, the decision on which components to adapt may not be unanimous. The process of updating these components dynamically involves transmitting a *configuration transaction* that all network members must endorse and verify according to a *system configuration endorsement policy*. To prevent the exclusion of any crucial decision-makers, a rigid endorsement policy must be established, and a manual override option must be implemented to reject any dynamic adaptation changes in the event of real-world conflicts among participants.

#### 8.1.3 Database

Two pluggable database options are available in Fabric – LevelDB and CouchDB. LevelDB uses an LSM tree, which is suitable for write-heavy workloads, while B-Trees are better for read-heavy workloads [34]. By utilizing the concepts from the literature, a self-driving LevelDB that can switch between the two architectures would be perfect for an evolving workload [34]. Similarly, there are studies on automated schema design and parameter tuning for NoSQL databases, which can be used to develop a self-driving CouchDB [35, 48, 50]. Designing a self-driving database is a complex research problem on its own. The challenge multiplies when designing a self-driving blockchain alongside a self-driving database. Nonetheless, as the database community has already made significant progress in this field [39, 40, 45–47, 57, 58, 76], there is a possibility of having a self-driving database that could be seamlessly integrated with a self-driving blockchain.

*8.1.4 Business Process Model*
Permissioned blockchains are mainly used in enterprise settings where the execution of applications is typically based on a business process model. Studies indicate that the business process model is closely linked with the performance of the underlying system, and several optimization strategies can improve the system throughput [10, 27]. The activities in a business process model correspond to the transactions in a blockchain system. Therefore, the effectiveness of these optimization strategies is dependent on the workload. Research is moving towards self-adapting business process models [13, 56], making them suitable candidates for an *adaptable feature*. Redesigning a business process model often means redesigning the smart contract, which requires authorization from all decision-making entities in a blockchain network. As a result, frequent updates may not be recommended. To address this issue multiple designs could be included in the initial business model, and these can be selected dynamically by the workflow engine.

## 8.2 Technology Independence

It is challenging to have a generic discussion for a self-driving blockchain system due to the vast implementation differences between different blockchain systems. Therefore, we use Fabric as an example to support our discussions in this paper. Our work can serve as a foundation to hold similar discussions about other blockchain platforms. For instance, Corda and Multichain, which are two other popular permissioned blockchains, have more than 9 and 15 dynamically adaptable system parameters, respectively [18, 19, 52]. Corda utilizes notaries to endorse transactions, and distributing transactions across multiple notaries is expected to enhance throughput, similar to Fabric's endorsement policies [17]. Quorum's (another popular permissioned blockchain) mining frequency, or block time, affects transaction latencies in a linearly proportional manner, similar to Fabric's block size [4]. Additionally, optimization strategies for Solidity smart contracts, which are used by many different blockchain systems, are extensively discussed in the literature [1, 8, 53] and can be potential candidates for dynamic adaptation. Finally, fairness is a universal concept that can be applied to all blockchain systems [32, 44]. Therefore, the client transaction rate is an adaptable feature independent of the blockchain platform. Our paper highlights the need for a self-driving blockchain and demonstrates its feasibility on Fabric. Although it may not be viable to reuse the same model on other blockchains, the methodology we have presented can be applied to other blockchain platforms.

## 8.3 Limitations

In our experiments, we opted for the DQN learning strategy, which is a widely-used approach in self-driving systems [6, 54, 59, 61]. However, there are several other learning strategies that we could have considered to improve the performance of our approach. The literature mentions alternative approaches such as recurrent neural networks, linear regression, actor-critic model, and deep deterministic policy gradient for designing self-driving systems [43, 46, 57, 76]. Further, in our experiments, we use a single performance metric to define our reward functions to reduce the time and complexity of our learning approach. However, several optimized approaches to defining the reward function with multiple performance metrics can be found in the literature [22, 29, 49].

Nevertheless, our primary objective is to provide a blockchain perspective to the discussions on self-driving systems. As a result, we aim to identify adaptable features in blockchains and demonstrate their feasibility. Our focus is not to compare and determine the perfect learning approach for self-driving systems since this topic has already been extensively discussed in the literature [3, 43, 45].

## 9 Related Work

The database community has conducted extensive research on self-driving, self-managing, auto-tuning and self-adaptive database management systems that target several areas, such as resource allocation, configuration parameters, query optimization, partitioning, storage layout, and indexes [11, 20, 31, 33, 39–41, 45–47, 57, 58, 75, 76]. However, our conversation is centred around distributed ledger technologies that differ in architecture and use cases. For instance, the number and type of configuration parameters available for tuning, as well as the presence of smart contracts, set blockchains apart from databases. Additionally, our discussion also explores self-driving possibilities at the application level in terms of fairness, which is more significant for decentralized systems such as blockchains.

In the recent literature, there have been discussions on the subject of self-adaptive and auto-tuning blockchains. For instance, Adachain [72] is a self-adaptive blockchain that modifies its architecture according to the incoming workload to enhance performance. Contrastingly, we focus on applying self-driving strategies to existing systems without modifying their core architecture, which helps users of established blockchain systems to employ our methodology without the need to switch to a new blockchain platform. Athena [43] is an auto-tuning system that can tune the configuration parameters of a blockchain before deployment. Our focus, on the other hand, is on parameter tuning of a live network. Sabine [42] is a self-adaptive blockchain that adapts the number of validators in its consensus protocol, while Ursa [62] can adjust the number of transactions in a block based on user requirements. These works emphasize a single adaptable aspect of blockchains, whereas our paper focuses on identifying adaptable features throughout the entire blockchain stack. Further, self-driving possibilites at the application level have not yet been explored in the literature.

## 10 Conclusions

The demand for self-driving blockchain systems is growing due to the increasing complexity and cost of maintaining existing blockchain applications. While some initial steps have been taken towards creating self-adaptive and auto-tuning blockchain systems, a comprehensive self-driving blockchain has yet to be explored. This paper focused on the opportunities for self-drive in Hyperledger Fabric, one of the most popular permissioned blockchains used by enterprises. Our investigation identified adaptable features at different levels of the blockchain stack that can be dynamically tuned to improve performance. We also addressed specific challenges and possible solutions. We set up three demonstrative autonomous systems and conducted extensive experiments to evaluate the feasibility of our findings. The results suggest that self-driving blockchain systems are a promising avenue for future research.